\documentclass[%
 reprint,
 amsmath,amssymb,superscriptaddress,
 aps,
pra,
]{revtex4-2}

\usepackage{physics}
\usepackage{graphicx}
\usepackage{dcolumn}
\usepackage{bm}
\usepackage{ulem}
\usepackage[dvipsnames]{xcolor}
\usepackage[colorlinks,allcolors=blue]{hyperref}

\newcommand{\one}{\mbox{$1 \hspace{-1.0mm}  {\bf l}$}}

\begin{document}

\preprint{APS/123-QED}

\title{Critical assessment of information back-flow in measurement-free teleportation}

\author{Hannah McAleese}
\affiliation{Centre for Quantum Materials and Technologies, School of Mathematics and Physics, Queen's University Belfast, BT7 1NN, United Kingdom}


\author{Mauro Paternostro}
\affiliation{Universit\`a degli Studi di Palermo, Dipartimento di Fisica e Chimica - Emilio Segr\`e, via Archirafi 36, I-90123 Palermo, Italy}
\affiliation{Centre for Quantum Materials and Technologies, School of Mathematics and Physics, Queen's University Belfast, BT7 1NN, United Kingdom}

\date{\today}

\begin{abstract}
We assess a scheme for measurement-free quantum teleportation from the perspective of the resources underpinning its performance. In particular, we focus on recently made claims about the crucial role played by the degree of non-Markovianity of the dynamics of the information carrier whose state we aim to teleport. We prove that any link between efficiency of teleportation and back-flow of information depends fundamentally on the way the various operations entailed by the measurement-free teleportation protocol are implemented, while -- in general -- no claim of causal link can be made. Our result reinforces the need for the explicit assessment of the underlying physical platform when assessing the performance and resources for a given quantum protocol and the need for a rigorous quantum resource theory of non-Markovianity. 
\end{abstract}

\maketitle


\section{Introduction}

Quantum teleportation~\cite{Bennett1993} shows the power of entanglement as a resource: by jointly measuring the quantum states of two particles, we can transfer, without any actual exchange of matter, a quantum state to a remote station. This process is commonly referenced in the context of quantum communication over long distances, but its applications to quantum computation are also paramount as evidenced by the success of the measurement-based model for quantum computing~\cite{Briegel2009}. The measurement-free teleportation protocol, put forward in Ref.~\cite{Brassard1998}, helped illustrate the centrality of entanglement by removing entirely measurements that, in contrast, are very important for the success of the original scheme~\cite{Bennett1993}.  

Ref.~\cite{Tserkis2022} challenged the nearly dogmatic view on the essential role of entanglement to explore the relation between the efficiency of measurement-free teleportation and non-Markovianity~\cite{deVega2017,BreuerReview2016,Rivas2014}. Specifically, the analysis by Tserkis {\it et al.} drew links between the information back-flow from the {\it instrumental part} of the computational register (considered as an environment) to its {\it relevant part} (the system), and the entanglement present in the environment. It is worth noticing that the original teleportation protocol has previously been studied in the context of non-Markovianity~\cite{Laine2014,Liu2020,Hesabi2021,Wang2023}, but such an assessment has normally been done by introducing an external environment. A different take to the {\it role} played by non-Markovianity in teleportation was addressed in Ref.~\cite{Laine2014}, where non-Markovianity was seen as an additive to performance rather than the mechanism underpinning it. Ref.~\cite{Wang2023}, instead, studied the use of non-Markovianity to mitigate against the effects of noise on the resource state.

This paper issues from the work of Tserkis~\cite{Tserkis2022} and assesses critically the link between non-Markovianity and efficiency in the measurement-free teleportation protocol. Methodologically, we model the teleportation circuit {\it as a quantum channel} for a system of interest~\cite{Bowen2001,Gu2004,Nielsen1997,Pirandola2017,Pirandola2018,Tserkis2018} and analyze the dynamics inherent within it.  

While we do not introduce non-Markovianity through any external means, by focusing on the measurement-free teleportation approach, we are in a position to  gain insight on the underlying dynamics of teleportation and analyze the non-Markovianity which could be inherently present in the protocol.

We show that such connections are -- at best -- very weak, and delicately dependent on the way the dynamics underpinning the protocol is implemented and interpreted. Only a very fine-grained assessment of the various stages of the teleportation channel allows us to unveil how non-Markovianity enters the dynamics of the register and, potentially, could play a role in the establishment of the right fluxes of information from the instrumental part of the teleportation register to its relevant part. On one hand, our results point towards the careful assessment of the way a dynamical map is implemented, in all of its sub-parts, before any conclusions on what embodies a {\it resource} of it can be made. On the other hand, it indirectly points to the need for a deeper understanding of the role played by non-Markovianity in quantum information problems and, in turn, the benefits of developing a comprehensive quantum resource theory of non-Markovianity~\cite{Berk2021,Bhattacharya2021}.

The remainder of this paper is organized as follows: in Sec.~\ref{sec:bgTeleportation} we illustrate the measurement-free teleportation protocol, and address it from the perspective of open-system dynamics, including the effects of its {\it channel description} on distinguishability of input states. Sec.~\ref{analysis} reviews the key instruments for our quantitative assessment of non-Markovianity and applies them to the evaluation of the information back-flow entailed by the measurement-free teleportation protocol. Sec.~\ref{corre} assesses in detail the link with quantum correlations shared by the relevant and ancillary part of the register, highlighting the controversial nature of claims linking such physical quantities and the performance of the scheme itself. Finally, Sec.~\ref{conc} offers our conclusions and perspectives. 

\section{Measurement-free teleportation} \label{sec:bgTeleportation}

\begin{figure}[b!]
    \centering
    \includegraphics[width=0.99\columnwidth]{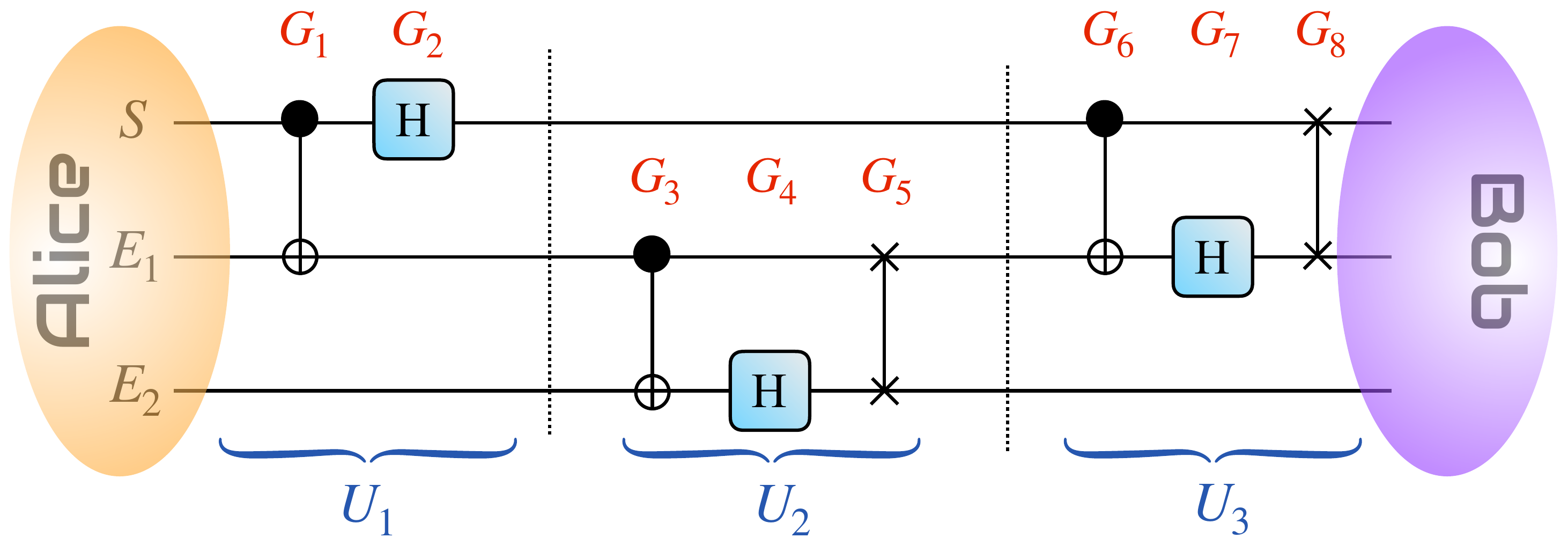}
    \caption{Quantum circuit of the measurement-free teleportation protocol. Each gate into which the circuit is decomposed is labelled as $G_i$ ($i\in\{1,\dots,8\}$). They can be grouped into three unitary {\it blocks} of operations $U_j~(j=1,2,3)$ as defined in Eqs.~\eqref{eq:U1}-\eqref{eq:U3}. Here, $G_{1,3,6}$ are CNOT gates, $G_{2,4,7}$ are Hadamard transforms, while $G_{5,8}$ are SWAP gates.}
    \label{fig:circuit}
\end{figure}

We follow the three-qubit measurement-free teleportation protocol put forward in Ref.~\cite{Brassard1998}, whose quantum circuit we present in Fig.~\ref{fig:circuit}. As our aim is to study the information back-flow and non-Markovianity in the protocol itself, it is appropriate to use the language of open quantum systems when describing it. We thus use the label $S$ for the {\it system} whose state is being teleported, and $E_{1,2}$ for the {\it environmental} particles that are ancillary for the protocol. 

Two agents, conventionally identified as Alice and Bob,  hold control of the full register consisting of system and environment as per Fig.~\ref{fig:circuit}. Alice aims to teleport to Bob the state $\ket{\psi(\alpha)} = \alpha \ket{0} + \sqrt{1-\vert\alpha\vert^2} \ket{1}$, which she encodes in qubit $S$. On the other hand, qubits $E_{1,2}$ make up a resource state. 

In the original formulation of the measurement-free protocol in Ref.~\cite{Brassard1998}, the teleported state was encoded in the degrees of freedom of one of the environmental qubits (specifically, $E_2$). The authors of Ref.~\cite{Tserkis2022} proposed an altered version of the scheme where $\ket{\psi}$ is recovered from qubit $S$. This, as mentioned before, put them in a position to argue for a direct relationship between the quality of retrieval of $\ket{\psi}$ from the degrees of freedom of $S$ and the non-Markovian character of the system-environment dynamics. As our scope is to critically assess the actual implications of non-Markovianity for the effectiveness of the protocol, we will adhere to the formulation in Ref.~\cite{Tserkis2022}. We emphasise that, though this modification is beneficial for studying and understanding the underlying dynamics, we would use the original formulation in Ref.~\cite{Brassard1998} to teleport states in a quantum computation context.   

In an ideal teleportation scheme, the resource encoded in the $E_1$-$E_2$ compound would be a maximally entangled Bell state. Here, in order to study the necessary correlations for the protocol at hand, we weaken this strong requirement and take the resource to be the Werner state
\begin{equation}
\label{eq:werner}
    W_{E_1E_2}(p) = p \ket{\phi^+}\bra{\phi^+}_{E_1E_2} + \frac{1-p}{4} \one_{E_1E_2} ,
\end{equation}
where $p\in[0,1]$ and $\ket{\phi^+}_{E_1E_2} = (\ket{00} + \ket{11})_{E_1E_2}/\sqrt{2}$ is a Bell state. Unless $p=0$, there are correlations present in $W_{E_1E_2}(p)$: it is entangled for $p\in(1/3,1]$ and carries quantum discord and classical correlations for $0 < p \leq 1$.

Alice sends $S$ and $E_1$ through the circuit to Bob. The operations $G_j~(j=1,...,8)$ undergone by the $S$-$E_1$ compound can be grouped into three {\it blocks} to highlight their roles in the process. The first is
\begin{equation} \label{eq:U1}
    U_1 = (\mathrm{H}_S \otimes \one_{E_1 E_2})(\mathrm{CNOT}_{SE_1} \otimes \one_{E_2}).
\end{equation}
Here $H=(X+Z)/\sqrt{2}$ stands for the Hadamard gate and $\mathrm{CNOT}_{SE_1}=\ket{0}\bra{0}_{S}\otimes\one_{E_1}+\ket{1}\bra{1}_{S}\otimes X_{E_1}$ is a controlled-NOT gate (with $X$ and $Z$ the Pauli $x$ and $z$ matrix, respectively). 
Owing to the interaction entailed by such a gate, quantum correlations might be established between $S$ and $E_1$ through the action of $U_1$. At this point, at least some of the information about the system state is encoded in the form of system-environment correlations. The degree with which this happens, though, depends on the initial state of the system: for $\alpha=0$, $S$  and the environment remain uncorrelated, while the total correlations are maximised for $\alpha=1/\sqrt{2}$. At this stage, due to the initial correlations within the environment, all elements of the register would be quantum correlated, in general. 

The second block takes the form of the unitary gate
\begin{equation} \label{eq:U2}
    U_2 = (\one_S \otimes \mathrm{SWAP}_{E_1 E_2}) (\one_{SE_1} \otimes \mathrm{H}_{E_2}) (\one_S \otimes \mathrm{CNOT}_{E_1 E_2})
\end{equation}
with $\mathrm{SWAP}_{E_1 E_2}$ the swap gate which, for any state $\ket{\phi}_{E_1}\ket{\eta}_{E_2}$, acts as 
\begin{equation}
    \mathrm{SWAP}_{E_1 E_2}\ket{\phi}_{E_1}\ket{\eta}_{E_2}=\ket{\eta}_{E_1}\ket{\phi}_{E_2}.
\end{equation}
In the ideal case where the environment is initially maximally entangled (i.e. for $p=1$), the $U_2$ operation acts to decouple $S$ and $E_2$, thereby localising the information encoded through $U_1$ between $S$ and $E_1$ only. Some correlations do remain between all three systems for $p<1$.

The final block of unitaries of the protocol is
\begin{equation} \label{eq:U3}
    U_3 = (\mathrm{SWAP}_{SE_1} \otimes \one_{E_2}) (\one_S \otimes \mathrm{H}_{E_1} \otimes \one_{E_2}) (\mathrm{CNOT}_{SE_1} \otimes \one_{E_2}).
\end{equation}

\subsection{Effective depolarising-channel description}

In the original protocol with $p=1$~\cite{Brassard1998}, all system-environment correlations vanish after application of $U_3$, and all the information about the input state is localised in the desired system. In the version discussed here, with an imperfect resource, the success of the protocol grows with $p$. 

In order to see this quantitatively, we resort to an effective description of the dynamics undergone by system $S$ as a result of the action of the quantum circuit and its coupling to the environmental qubits. We call $\rho^i_S=\ket{\psi}\bra{\psi}_S$ the initial state of the system qubit, label as ${\cal U}=U_3U_2U_1$ the total unitary of the circuit and decompose the identity $\openone_{E_1E_2}$ in the Hilbert space of the environmental compound over the Bell basis $\{\ket{{\cal B}^\alpha_j}\}_{E_1E_2}\equiv\{\ket{\phi^\pm},\ket{\psi^\pm}\}_{E_1E_2}$, where $\alpha=\phi,\psi$ and $j=\pm$ and we introduce the remaining Bell states
 \begin{equation}
 \begin{aligned}
     \ket{\phi^-}_{E_1E_2}&=\frac{1}{\sqrt2}\left(\ket{00}-\ket{11}\right)_{E_1E_2},\\
     \ket{\psi^\pm}_{E_1E_2}&=\frac{1}{\sqrt2}\left(\ket{01}\pm\ket{10}\right)_{E_1E_2}.
     \end{aligned}
 \end{equation}
 The final state of $S$ thus reads
\begin{equation}
\label{eq:finalS}
\rho^f_S=\text{Tr}_{E_1E_2}\left[{\cal U}\left(\rho^i_S\otimes W_{E_1E_2}(p)\right){\cal U}^\dag\right].    
\end{equation} 
 Upon inserting Eq.~\eqref{eq:werner} into this expression, we have
 \begin{equation}
 \label{eq:explicit}
 \begin{aligned}
          &\rho^f_S=p\!\!\sum_{j,k=0,1}{}_{E_1E_2}\!\bra{jk}{{\cal U}}\ket{\phi^+}_{E_1E_2}\rho^i_S{}_{E_1E_2}\!\bra{\phi^+}{{\cal U}^\dag}\ket{jk}_{E_1E_2}\\
          &+\frac{1-p}{4}\!\!\sum_{j,k,\alpha,l}{}_{E_1E_2}\!\bra{jk}{{\cal U}}\ket{{\cal B}^\alpha_l}_{E_1E_2}\rho^i_S{}_{E_1E_2}\!\bra{{\cal B}^\alpha_l}{{\cal U}^\dag}\ket{jk}_{E_1E_2} \\  
          &=\frac{1+3p}{4}\!\sum_{j,k=0,1}{{\cal V}}^{\phi^+}_{jk}\rho^i_S{{\cal V}}^{\phi^+\dag}_{jk}+\frac{1-p}{4}\sum_{j,k,\alpha,l}' {{\cal V}}^{\alpha^l}_{jk}\rho^i_S{{\cal V}}^{\alpha^l\dag}_{jk},
          \end{aligned}     
 \end{equation}
where the symbol ${\sum}'_{\alpha,l}$
stands for the summation over all the elements of the Bell basis except $\ket{\phi^+}_{E_1E_2}$ and we have introduced the operators ${\cal V}^{\alpha^l}_{jk}={}_{E_1E_2}\bra{jk}{{\cal U}}\ket{{\cal B}^\alpha_l}_{E_1E_2}$ of the open-system dynamics undergone by $S$. An explicit calculation leads to the results summarized in Table~\ref{tab1}.
\begin{table}[b]
    \centering
    \begin{tabular}{c|c|c|c|c}
    \textbf{${\cal V}^{\alpha^l}_{jk}$} & $j=0,k=0$ & $j=0,k=1$ & $j=1,k=0$ & $j=1,k=1$\\
    \hline
    \hline
    ${\cal V}^{\phi^+}_{jk}$ & $\openone_S/2$ & $\openone_S/2$ & $\openone_S/2$ &$\openone_S/2$\\
    \hline
    ${\cal V}^{\phi^-}_{jk}$ & $Z_S/2$ & $-Z_S/2$ & $Z_S/2$ & $-Z_S/2$\\
    \hline
    ${\cal V}^{\psi^+}_{jk}$ & $X_S/2$ & $X_S/2$ & $-X_S/2$ & $-X_S/2$\\
    \hline
    ${\cal V}^{\psi^-}_{jk}$ & $-iY_S/2$ & $iY_S/2$ & $iY_S/2$ & $-iY_S/2$\\
    \hline
    \end{tabular}
    \caption{Summary of the explicit form taken by the operators ${\cal V}^{\alpha^l}_{jk}$ acting in the Hilbert space of $S$ [cf. Eq.~\eqref{eq:explicit}].}
    \label{tab1}
\end{table}
Using such expressions, we are finally able to recast the final state of the system in the form of the operator-sum decomposition $\rho^f_S=\sum^4_{j=1}{\cal K}_{j}\rho^i_S{\cal K}^\dag_j$ with
\begin{equation}
\begin{aligned}
{\cal K}_1&=\sqrt{\frac{1+3p}{4}}\openone_S,\quad{\cal K}_2=\sqrt{\frac{1-p}{4}}Z_S,\\
{\cal K}_3&=\sqrt{\frac{1-p}{4}}X_S,\quad{\cal K}_4=\sqrt{\frac{1-p}{4}}Y_S,
\end{aligned}   
\end{equation}
which immediately give $\sum^4_{j=1}{\cal K}^\dag_j{{\cal K}}_j=\openone_S$ and allows us to conclude that the action of the measurement-free teleportation protocol on the state of the system is that of a depolarising channel acting with a resource-dependent rate $q=1-p$. The corresponding state fidelity with $\rho^i_S$ reads 
\begin{equation}
F(p){=}_S\!\bra{\psi}\rho^f_S\ket{\psi}_S=(1+p)/2,
\end{equation}
thus increasing linearly from $1/2$ when $p=0$, to 1 when $p=1$. The role of the $\mathrm{SWAP}_{SE_1}$ gate in $U_3$ is to transfer the information on $\ket{\psi}$ otherwise encoded in the state of $E_1$ to the system qubit $S$. As already anticipated, the inclusion of this gate in the protocol allows us to characterize the quality of the teleportation performance in terms of state-revival in the system qubit.

\subsection{Distinguishability and non-Markovianity resulting from the dynamics} \label{sec:measures}

While this analysis shows  the non-trivial nature of the overall action of the quantum circuit on the state of $S$, it is instructive to dissect the effects of the individual $U_j$, particularly in terms of the degree of distinguishability of different input states of $S$. In order to do this in a quantitative manner, we make use of the instrument embodied by the 
trace distance between two quantum states. This is defined as 
\begin{equation}
    D(\rho_1,\rho_2) = \frac{1}{2} \| \rho_1 - \rho_2 \|_1,
\end{equation}
where $\rho_{1,2}$ are two arbitrary density matrices and  $\|A\|_1 = \mathrm{Tr}[\sqrt{A^\dagger A}]$ is the trace norm of an arbitrary matrix $A$. 

First, let us consider the action of $U_1$ on the initial state of $S$. Following an approach fully in line with the one formalized in Eq.~\eqref{eq:explicit} but for ${\cal U}\to U_1$ and by labelling the state of $S$ resulting from the application of this block of unitaries alone as $\rho^1_S(\alpha)$, so as to emphasize the dependence on the initial-state parameter $\alpha$, we have 
\begin{equation}
\label{eq:rho1S}
\rho^1_S(\alpha)=\sum_{i=1,2}K_i\rho^i_S K^\dag_i=\alpha^2\ket{+}\bra{+}_S+(1-\alpha^2)\ket{-}\bra{-}_S,  
\end{equation}
where we have introduced the Kraus operators $K_1=\ket{+}\bra{0}_S$ and $K_1=\ket{-}\bra{1}_S$, which are written in terms of the eigenstates $\ket{\pm}_S$ of $X_S$ such that $X_S\ket{\pm}_S=\pm\ket{\pm}_S$. We thus consider 
\begin{equation}
    D(\rho^1_S(\alpha_1),\rho^1_S(\alpha_2))=|\alpha^2_1-\alpha^2_2|,
\end{equation}
where $\alpha_{1,2}\in\mathbb{R}$ (without loss of generality) identify two different initial states of the system. Having in mind the analysis of the degree of non-Markovianity that will be presented later on in this work, we take $\alpha_1=1$ and $\alpha_2=0$ (so as to prepare $S$ in eigenstates of $Z_S$), and thus consider fully distinguishable input states. For such a choice, we have $D(\rho^1_S(\alpha_1),\rho^1_S(\alpha_2))=1$, achieving again full distinguishability regardless of the properties of the environmental system (as $\rho^1_S(\alpha)$ does not depend on $p$).

As for $U_2$, it is clear from Fig.~\ref{fig:circuit} that this block of unitaries is {\it local} with respect to the bipartition $S$-$(E_1E_2)$. That is, $U_2$ does not contain degrees of freedom of $S$, which implies that the corresponding operator-sum decomposition of the effective channel acting on the system involves only the identity operator $\openone_S$. The evolved state $\rho^2_S$ after $U_2$ is thus identical to Eq.~\eqref{eq:rho1S}. Notice, though, that the state of the environment will be changed by this part of the circuit. 

Finally, block $U_3$ will need to be applied to the -- in general quantum correlated -- joint state of $S$-$(E_1E_2)$. This immediately gives evidence of the fundamental difference between the action entailed by $U_3$ and the other blocks of operations: while, as for $U_1$, this operation couples $S$ to the environment, the input state to $U_3$ is a state that features, as mentioned above, system-environment correlations that {\it may} play a key role in determining the nature of the dynamics of $S$. Technically, such correlations prevent us from using the same approach as above to identify the effective channel acting on $S$. Instead, we will have to calculate
\begin{equation}
\begin{aligned}
    \rho^3_S&\equiv\rho^f_S=\Phi_{p}(\rho^i_S)={\text Tr}_{E_1E_2}\left[U_3\rho^2_{SE_1E_2}U_3^\dag\right]\\
    &=\frac12
\begin{pmatrix}
 2 \alpha ^2p-p+1 & 2\alpha  \sqrt{1-\alpha ^2} p \\
 2\alpha  \sqrt{1-\alpha ^2} p & -2 \alpha ^2p +p+1 
\end{pmatrix}
\end{aligned}
\end{equation}
with $\rho^2_{SE_1E_2}$ the output state of the system-environment compound after application of $U_2U_1$, and $\Phi_p(\cdot)$ the $p$-dependent dynamical map resulting from taking the trace over the environmental degrees of freedom. The trace distance between two input states of $S$ reads
\begin{equation}
\begin{aligned}
    D&(\rho^f_S(\alpha_1),\rho^1_f(\alpha_2))=p\left\vert\alpha_1\sqrt{1-\alpha^2_2}-\alpha_2\sqrt{1-\alpha^2_1}\right\vert\\
    &=pD(\ket{\psi(\alpha_1)}\bra{\psi(\alpha_1)}_S,\ket{\psi(\alpha_2)}\bra{\psi(\alpha_2)}_S).
    \end{aligned}
\end{equation}
This shows that the last block of the quantum circuit at hand is the only one that could change the degree of distinguishability between the input state and evolved one, which in general shrinks linearly with the depolarisation rate. 

\section{Analysis of non-Markovianity in the measurement-free teleportation circuit}
\label{analysis}

We are now in a position to leverage the tools and results achieved so far to assess the non-Markovian features of the dynamics entailed by the measurement-free teleportation protocol. We start by briefly reviewing some of the most popular measures of non-Markovianity reported so far in literature. 

\subsection{Review of measures of non-Markovianity}
\label{revNM}

In this short review, we do not aim to be comprehensive and refer the interested reader to Refs.~\cite{Rivas2014,deVega2017,BreuerReview2016}. 

\subsubsection{Breuer-Laine-Piilo measure}

The measure of non-Markovianity which we will rely on the most is the one proposed by Breuer, Laine and Piilo (BLP) in Ref.~\cite{Breuer2009}. It builds on the fact that the trace distance decreases monotonically under  Markovian dynamics, any increase signalling non-Markovianity. 

The measure is therefore calculated as
\begin{equation}
    \mathcal{N}_{BLP}= \max_{\rho_{1,2} (0)} \int_{\sigma >0} \sigma(t, \rho_{1,2}(0))\,dt,
\end{equation}
where $\sigma (t,\rho_{1,2} (0)) = \partial_tD(\rho_1 (t),\rho_2 (t))$ is the rate of change of the trace distance between two initial states $\rho_{1,2}(0)$ of the systems we are hoping to distinguish, considered over all the time intervals $(a_i,b_i)$ where $\sigma >0$ (i.e. where the trace distance increases). This can alternatively be expressed in a simpler form as 
\begin{equation}
    \mathcal{N}_{BLP} = \max_{\rho_{1,2} (0)} \sum_i [ D  (\rho_1(b_i), \rho_2 (b_i))  - D(\rho_1(a_i),\rho_2(a_i)) ].
\end{equation}
Note that growth in trace distance is a necessary but not sufficient condition for non-Markovianity. We therefore include other measures in this paper to ensure that we are not missing non-Markovian dynamics in the protocol even when $\mathcal{N}_{BLP} = 0$.

We can simplify the optimisation procedure by using the result that optimal initial state pairs will be antipodal points on the surface of the Bloch sphere~\cite{Wissmann2012}.

\subsubsection{Rivas-Huelga-Plenio measure}

The Rivas-Huelga-Plenio (RHP) measure~\cite{Rivas2010}, on the other hand, is based on a necessary and sufficient condition for Markovianity. A dynamical map between two times $t+\epsilon$ and $t$ where $t>0$ can be written as
\begin{equation}
    \mathcal{E}_{(t+\epsilon,t)} = \mathcal{E}_{(t+\epsilon,0)} \mathcal{E}^{-1}_{(t,0)},
\end{equation}
where the inverse map $\mathcal{E}^{-1}_{(t,0)}$ may not be completely positive. Given a maximally entangled state of the system and an ancilla, such as $\ket{\phi^+}_{SA}$, we apply the dynamical map $\mathcal{E}_{(t+\epsilon,t)} \otimes \one$. The dynamics are Markovian if and only if $f_{NCP} = 1$ where 
\begin{equation}
    f_{NCP} (t+\epsilon,t) = \| (\mathcal{E}_{(t+\epsilon,t)} \otimes \one) \ket{\phi^+}\bra{\phi^+} \|_1,
\end{equation}
which can be recast as $g(t) = 0$ with
\begin{equation}
    g(t) = \lim_{\epsilon\rightarrow 0^+} \frac{f_{NCP} (t+\epsilon,t) -1}{\epsilon}.
\end{equation}
The corresponding measure of non-Markovianity is then
\begin{equation} \label{eq:measure}
    \mathcal{N}_{RHP} = \int_0^\infty g(t)\, dt.
\end{equation}
The quantity measure $\mathcal{N}_{RHP}/2$ is a lower bound of the robustness of non-Markovianity~\cite{Bhattacharya2021}. We can therefore attribute a physical meaning to it, that is, the amount of noise which must be added to a non-Markovian operation before it becomes Markovian.

\subsubsection{Luo-Fu-Song measure}

The correlations-based measure proposed by Luo, Fu and Song (LFS) in Ref.~\cite{Luo2012} follows a similar reasoning as the BLP one. The quantum mutual information between a system and an ancilla $I(\rho_{SA})$, which can be calculated as
\begin{equation}
    I(\rho_{SA}) = S(\rho_S) + S(\rho_A) - S(\rho_{SA}),
\end{equation}
where $S(\rho)=-\mathrm{Tr} (\rho \log_2 \rho)$ is the von Neumann entropy of state $\rho$, is monotonically decreasing under Markovian dynamics acting on the system only. Therefore, any increase in mutual information witnesses non-Markovianity.

We write the rate of change of the mutual information as $\gamma(\rho_{SA}(t)) = \partial_tI(\rho_{SA}(t))$ with 
$\rho_{SA}(t) = (\Phi_S \otimes \one_A) \rho_{SA}(0)$. The measure building on this rate is therefore given by
\begin{equation}
    \mathcal{N}_{LFS} = \int_{\gamma > 0} \gamma(\rho_{SA}(t))\,dt,
\end{equation}
where the integral is taken, once more, in the regions where $\gamma>0$. In line with the Choi-Jamio{\l}kowski isomorphism, we take the initial state $\rho_{SA}(0)$ to be any maximally entangled state to evaluate the measure. 

\subsection{Information back-flow and non-Markovianity} \label{sec:nonMarkovianity}

We begin by taking a straightforward approach to the description of the generator of the dynamical map. We provide an effective Hamiltonian of the system and environment by considering
\begin{equation} \label{eq:hamiltonian}
    H_\text{eff} = -i \ln{\cal U},
\end{equation}
where we have assumed units such that $\hbar=1$. Fixing a branch for the logarithm makes it single-valued, and $H$ uniquely determined from ${\cal U}$. The inspection of $H_\text{eff}$, which we do not report here as too complex, reveals that -- in general -- such a Hamiltonian involves three-body interactions between $S$, $E_1$ and $E_2$.

\begin{figure}
    \centering
    \includegraphics[width=0.95\columnwidth]{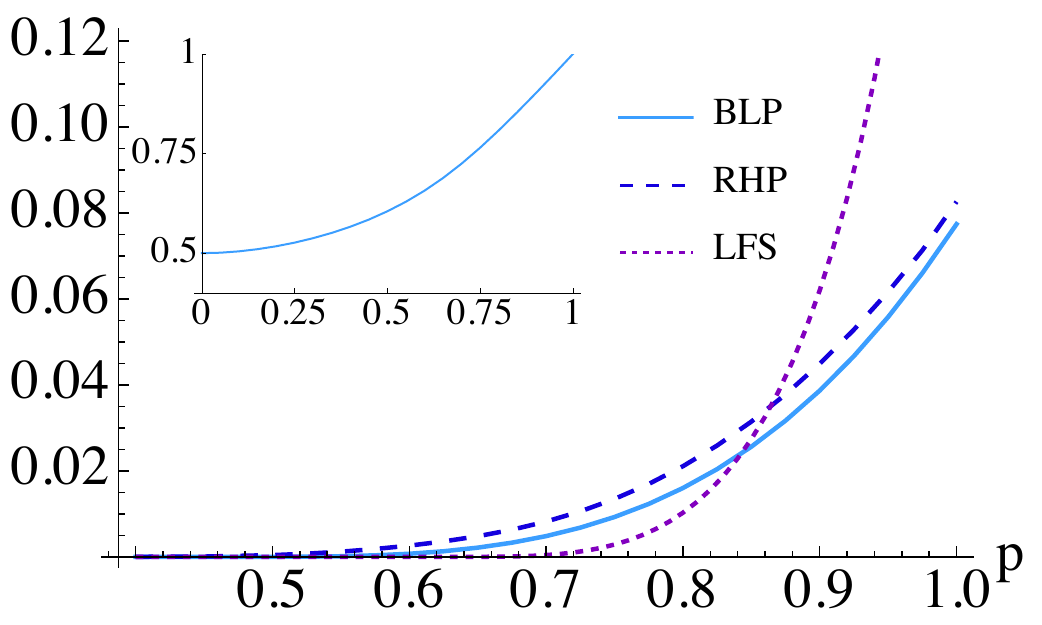}
    \caption{Non-Markovianity of the dynamics given by Eq.~\eqref{eq:hamiltonian}. Results are plotted only for $p\geq 0.4$ as all the measures listed in Sec.~\ref{revNM} are zero for $0\leq p \leq 0.4$. Inset: Non-Markovianity of the effective Hamiltonian in Eq.~\eqref{eq:8Hamiltonian} as measured by the ${\cal N}_{BLP}$ measure.}
    \label{fig:measuresNonM}
\end{figure}

Considering the corresponding time-evolution operator $e^{-i H_\text{eff}t}$ and varying the parameter $p$, we evaluate each of the measures reviewed in Sec.~\ref{revNM}. The results are plotted in Fig.~\ref{fig:measuresNonM}. The degree of non-Markovianity of the dynamics of $S$ remains small for each of the measures, and is virtually negligible for states of the environment with $p\lesssim 0.7$. However, as expected, the RHP measure is the most sensitive and detects non-Markovianity within a larger range of values of the mixing parameter in Eq.~\eqref{eq:werner}: the lowest value witnessing non-Markovianity through RHP is $p=0.41$, to be compared to $p=0.5$ for BLP and $p=0.65$ for LFS. As Werner states are inseparable for $p>1/3$, the environment is always entangled when the dynamics are non-Markovian. 

We can, however, take a different approach to the dynamics. Instead of working {\it block by block}, we use a fine-grained approach that considers each $G_j$ gate in sequence [cf. Fig.~\ref{fig:circuit}]. We thus address the eight effective Hamiltonian operators 
\begin{equation}~\label{eq:8Hamiltonian}
    H_{\text{eff},j}= - i \log G_j\quad (j=1,..,8)
\end{equation}
and use the corresponding time-evolution operators to describe the dynamics. This assumption results in dramatically different values of the BLP measure, which are displayed in Fig.~\ref{fig:measuresNonM}, thus proving that the unitary block-based approach was too coarse grained to gather the features of the circuit dynamics. Surprisingly, in this case we find non-Markovianity even when the environment is prepared in a maximally mixed state (i.e. for $p=0$). The BLP measure gives much larger values in this case compared to the previous modelling. While the initial states of $S$ that should be used in the calculation of the BLP measure in the block-based approach are the eigenstates of $Z_S$, in the individual gate-based model such states change depending on $p$. 

\begin{figure}[b!]
    \centering
    \includegraphics[width=0.8\columnwidth]{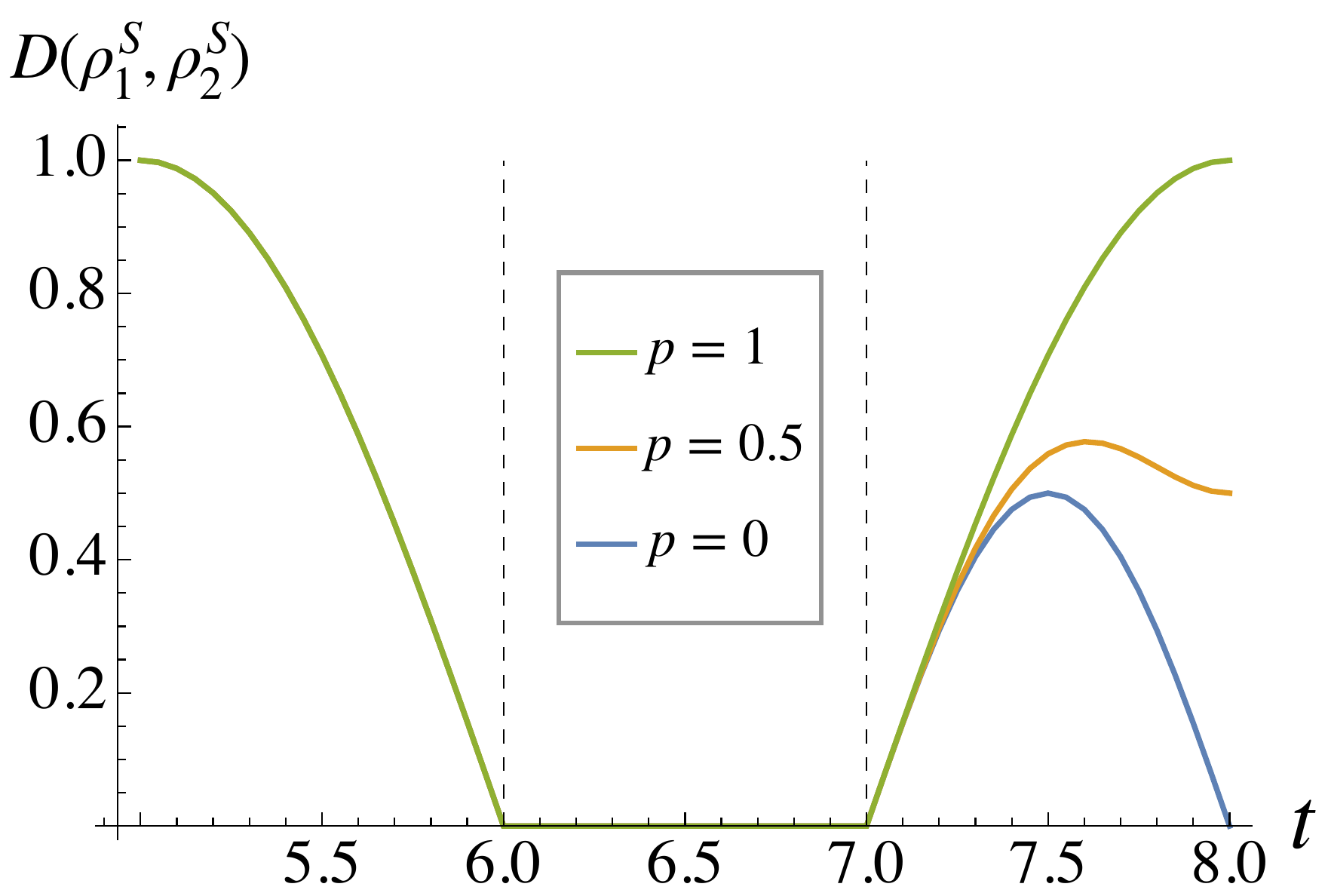}
    \caption{Trace distance between the two states of $S$ as they go through the teleportation circuit. Gate $G_i$ acts when $i-1 < t < i$. We plot only $5\leq t \leq 8$ as the trace distance remains constant at 1 for $t<5$.}
    \label{fig:traceDistanceS}
\end{figure}

\begin{figure}[t!]
    \centering
    \includegraphics[width=0.8\columnwidth]{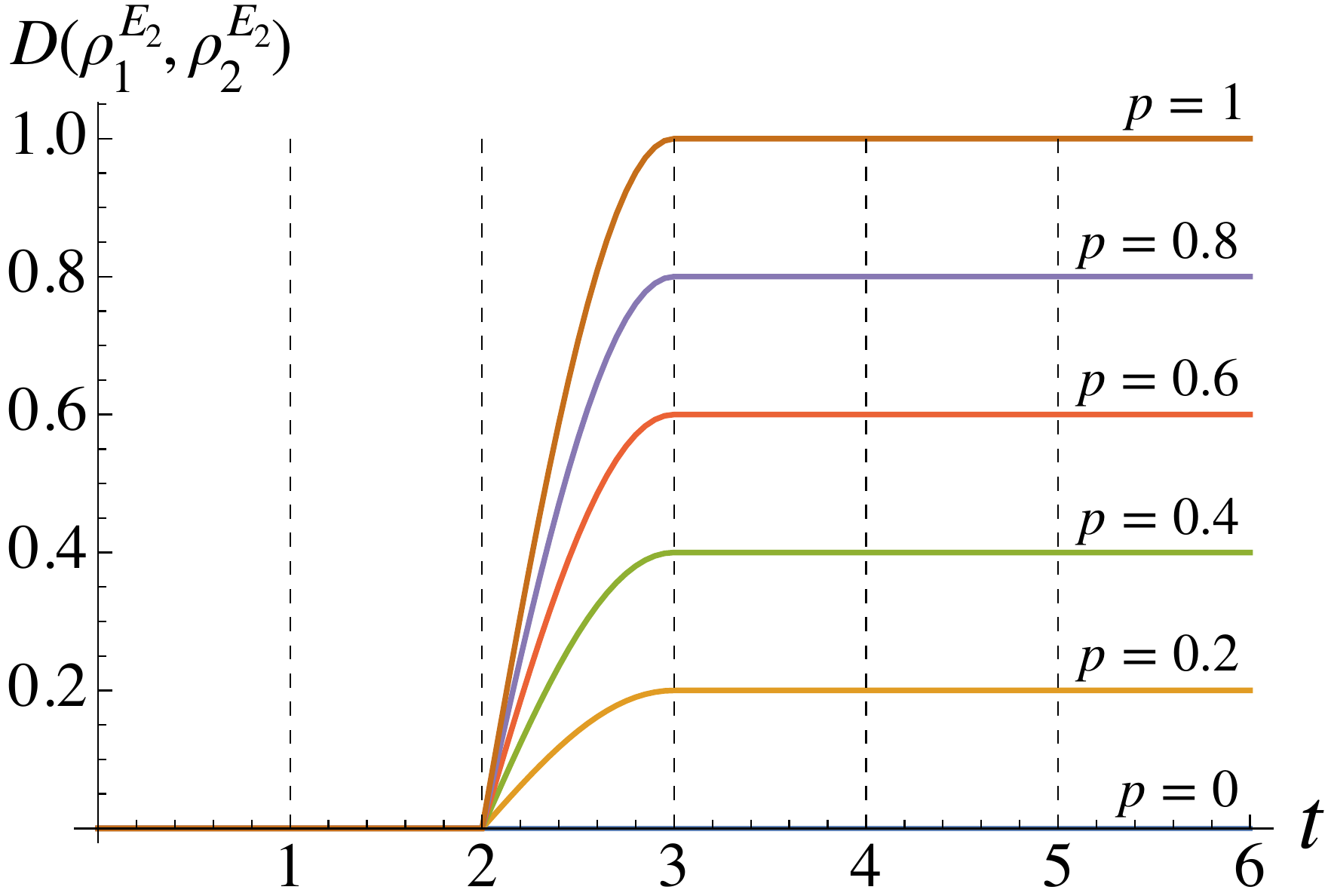}
    \caption{Trace distance between the two states of $E_2$ for the original BBC protocol~\cite{Brassard1998} as time evolves. The dashed lines are boundaries between the gates acting on the system and environment. Gates $G_1$-$G_4$ correspond to those in Fig.~\ref{fig:circuit}, while $G_5$ is a CNOT operation on $SE_2$ and $G_6$ is a Hadamard gate on $E_2$.}
    \label{fig:traceDistanceE2}
\end{figure}

What is the physical origin of the non-null degree of  non-Markovianity for $p=0$? In Fig.~\ref{fig:traceDistanceS}, we plot the trace distance when $p=0$ and we begin with $\ket{0}$ and $\ket{1}$ in $S$ (the optimal states when the environment is completely uncorrelated). We see that the trace distance increases only during the last gate, a SWAP operation between $S$ and $E_1$. This is true even when we change the initial state of the system. We remark that this gate is not in the original measurement-free teleportation circuit, but was added in Ref.~\cite{Tserkis2022} so as to consider the impact of the protocol on one single system. The original circuit starts with the same gates $G_1$-$G_4$ as in Fig.~\ref{fig:circuit}, followed by a CNOT between $S$ and $E_2$ and a Hadamard gate acting on $E_2$~\cite{Brassard1998}. The initial system state $\ket{\psi}$ is therefore teleported to $E_2$ without returning to $S$. If we revert back to the original protocol and consider the dynamics of $E_2$, tracing out $S$ and $E_1$, we can see if there is any increase in trace distance. This is similar to the method used in Ref.~\cite{Bowen2001}, for example. We have plotted the results in Fig.~\ref{fig:traceDistanceE2}.

We find that the trace distance remains zero for $p=0$, but does grow for $p>0$. In fact, in terms of the BLP measure we find that $\mathcal{N}_{BLP}  = p$. Therefore, only for $p=0$ can we claim that the non-Markovianity comes merely from the extra SWAP gate. Though the results in Fig.~\ref{fig:traceDistanceE2} are limited to the initial state pair $\{ \ket{0}, \ket{1} \}$, we find that there is no increase in trace distance for any alternative initial states. However, for $p>0$ we can see that trace distance does indeed increase and non-Markovianity is therefore present even when we label $E_2$ as our ``system" instead of $S$.

\section{Information back-flow and correlations}
\label{corre}

In the previous Section, we discovered that the relation between non-Markovianity, the performance of the teleportation scheme, and the entanglement in the initial state of the environment depends on whether  the implementation of the circuit allows for the consideration of the individual $G_j$ gates rather than the blocks of unitaries playing key roles in the evolution of the state of $S$. In the latter arrangement, the dynamics of $S$ is  non-Markovian for $p\geq 0.41$; in the former, non-Markovianity is present in the map evolving $S$ even when the environment is in a separable state, thus breaking the connection established in Ref.~\cite{Tserkis2022}. We now study the relation between non-Markovianity of the dynamics and system-environment correlations as time evolves. 

\begin{figure*}[t!]
    \centering
    \includegraphics[width=0.9\textwidth]{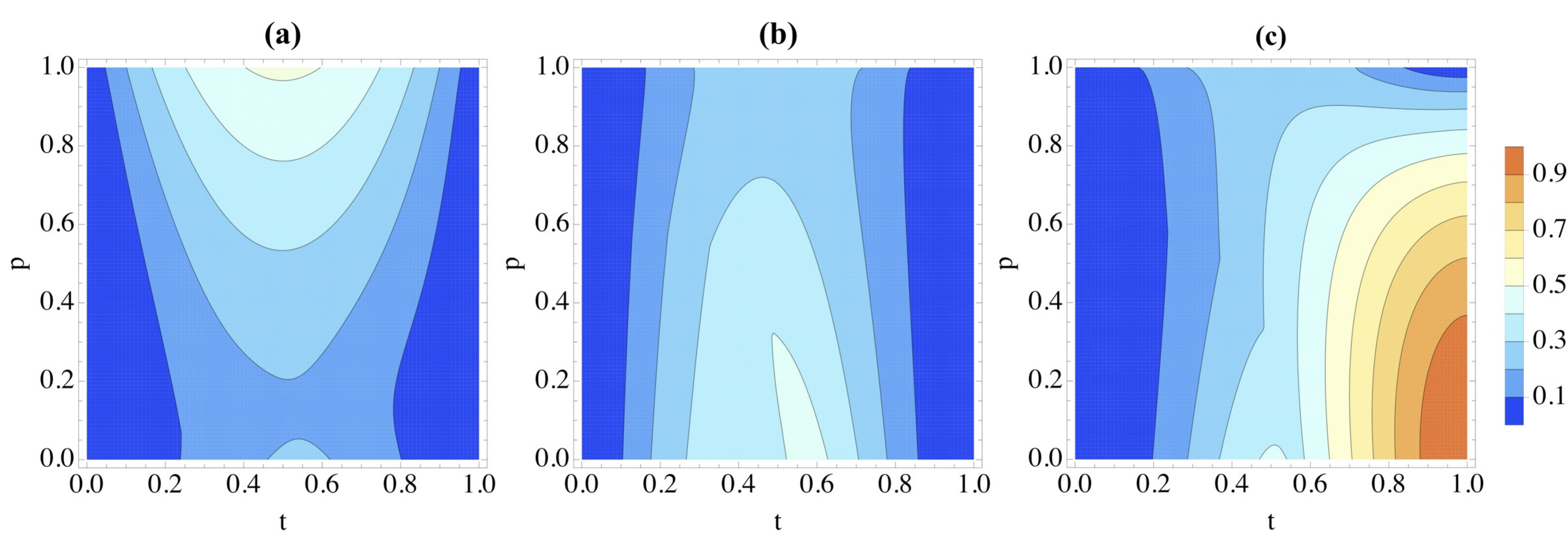}
    \caption{Correlations in the splitting $S$-vs-$(E_1E_2)$, as quantified by {\bf (a)} logarithmic negativity, {\bf (b)} discord and {\bf (c)} classical correlations as the system evolves according to Eq.~\eqref{eq:hamiltonian}. Time is denoted $t$ and $p$ determines the Werner state of the environment at $t=0$. The system is initially in the vacuum state $\ket{0}$.}
    \label{fig:sysEnvCorrOverlap}
\end{figure*}

Initially, correlations are only present in the environmental Werner state. As done previously,  we begin by describing the dynamics using the Hamiltonian in Eq.~\eqref{eq:hamiltonian}. Fig.~\ref{fig:sysEnvCorrOverlap} {\bf (a)} displays the entanglement between the system and environment as time evolves from an input state of $\ket{\psi}_S = \ket{0}$, as measured by the logarithmic negativity~\cite{Plenio2005,Horodecki2009}
\begin{equation}
  E = \log_2 \| \rho_{SE_1E_2}^{T_S} \|_1
\end{equation}
to quantify the entanglement in the bipartition $S$-vs-$(E_1E_2)$, where $\rho^{T_S}_{SE_1E_2}$ is the partial transpose of the evolved state $S$-$(E_1E_2)$ compound with respect to $S$. As might have been expected, the larger the initial environmental entanglement (as related to $p$), the more entanglement is shared between $S$ and $(E_1 E_2)$ during the protocol, and this corresponds to a larger degree of non-Markovianity. 

However, the growth in entanglement when $p\approx 0$ is quite surprising: $S$ and the environment are more entangled when $p=0$ than when such parameter takes a small ($\lesssim0.2$) yet non-zero value, even though the environment has more quantum and classical correlations, initially, in this case. This could relate also to correlations of a nature that are different from entanglement. In order to address this, we use figures of merit for quantum and classical correlations defined as in Refs.~\cite{ollivier2001,henderson2001}. First, we quantify classical correlations in a bipartite system composed of $A$ and $B$ using the generalised conditional entropy
\begin{equation}
    J(A|B) = \max_{B^\dagger_i B_i} \left( S(\rho_A) - \sum_{i} p_i S(\rho^i_A) \right),
\end{equation}
where $B^\dagger_i B_i$ is a POVM on system $B$, $\rho_A=\mathrm{Tr}_B [\rho_{AB}]$ and $\rho_A^i$ is the state of $\rho_A$ after system $B$ has been measured with $B^\dagger_i B_i$. This enables us to find the maximum information we can gain about system $A$ by measuring system $B$. As for quantum correlations, we resort to discord~\cite{Modi}, namely the difference between total correlations (as measured by the quantum mutual information) and classical correlations
\begin{equation}
    \mathcal{D} (A|B) = I(\rho_{AB}) - J(A|B).
\end{equation}
For simplicity, the maximum entailed by the definition of $J(A|B)$ will be sought over all projective measurements only, following the examples in Refs.~\cite{Galve2011,Al-qasimi2011}. While this is accurate and rigorous only for two-qubit systems, for our three-qubit problem we will only be able to quantify lower (upper) bounds to  classical (quantum) correlations.

Starting from the same initial state of $\ket{\psi}_S = \ket{0}$, discord and classical correlations for the bipartition $S$-vs-$(E_1E_2)$ are shown in Figs.~\ref{fig:sysEnvCorrOverlap} {\bf (b)} and {\bf (c)}, where we can appreciate a behavior that is, qualitatively, the inverse of entanglement: larger degrees of discord and  mutual information are found in the state at hand as $p$ decreases, which is somewhat counterintuitive. Therefore, while entanglement and non-Markovianity may be connected, we can conclude that discord and classical correlations are not linked to non-Markovianity. 

It is important to note that, at the end of the protocol (i.e. for $t=1$), only classical system-environment correlations remain for $p<1$. The information about $\ket{\psi}$ is encoded in such correlations and, thus, information back-flow is prevented. This is the reason behind the reduced success of the protocol as $p$ diminishes.

As the initial state of the system directly affects how entanglement is shared during the protocol (as highlighted in Sec.~\ref{sec:bgTeleportation}), without affecting the performance of the protocol, we addressed the case of inputting state $\ket{+}$ rather than $\ket{0}$. However, the results were similar with all the same features visible in the behavior of each figure of merit of correlations.  

\begin{figure*}[t!]
    \centering
    \includegraphics[width=0.9\textwidth]{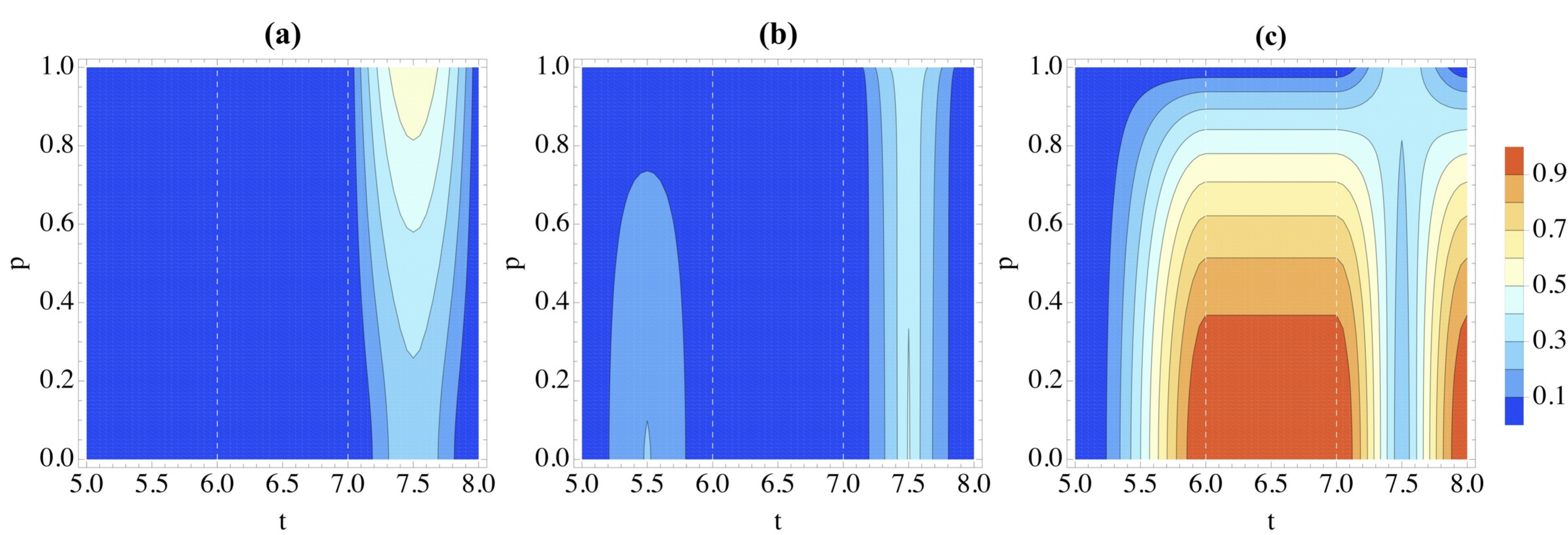}
    \caption{Correlations in the partition $S|E_1 E_2$ as quantified by {\bf (a)} logarithmic negativity, {\bf (b)} discord and {\bf (c)} classical correlations when the Hamiltonian of the system and environment is given by Eq.~\eqref{eq:8Hamiltonian}. We take the initial state of the system to be $\ket{0}$ and the environment $W(p)$. Here $t$ is a dimensionless time. We only show $5\leq t \leq 8$ as there are no system-environment correlations before $t=5$.}
    \label{fig:sysEnvCorrDistinct0}
\end{figure*}

As in Sec.~\ref{sec:nonMarkovianity}, we now change the dynamics to that in Eq.~\eqref{eq:8Hamiltonian}, and thus assume that each gate can be performed one-by-one in an independent way. We begin, as before, with the initial system state $\ket{\psi}_S = \ket{0}$. The correlations are shown in Fig.~\ref{fig:sysEnvCorrDistinct0}, which displays some similarities with the study performed in Fig.~\ref{fig:sysEnvCorrOverlap}. As in the previous case, the entanglement between $S$ and $E_1E_2$ is larger for larger $p$ [cf.~Fig.~\ref{fig:sysEnvCorrDistinct0} {\bf (a)}]. However, for this implementation of the quantum circuit operations, there is no unexpected growth in entanglement for $p\simeq0$. Moreover, entanglement only appears when the final gate of the circuit is performed, which is precisely when the trace distance rises in Fig.~\ref{fig:traceDistanceS} signalling non-Markovianity. This all heavily implies that entanglement is necessary for non-Markovian dynamics in the protocol.

The discord and classical correlations in Fig.~\ref{fig:sysEnvCorrDistinct0} {\bf (b)} and {\bf (c)} also share features of those in Fig.~\ref{fig:sysEnvCorrOverlap}; they are both larger for smaller $p$. When $p=1$, these correlations grow and vanish only during the course of $G_8$, the SWAP gate between $S$ and $E_1$. However, they can also appear during $G_6$ when $p<1$. 

\begin{figure*}[t!]
    \centering
    \includegraphics[width=0.9\textwidth]{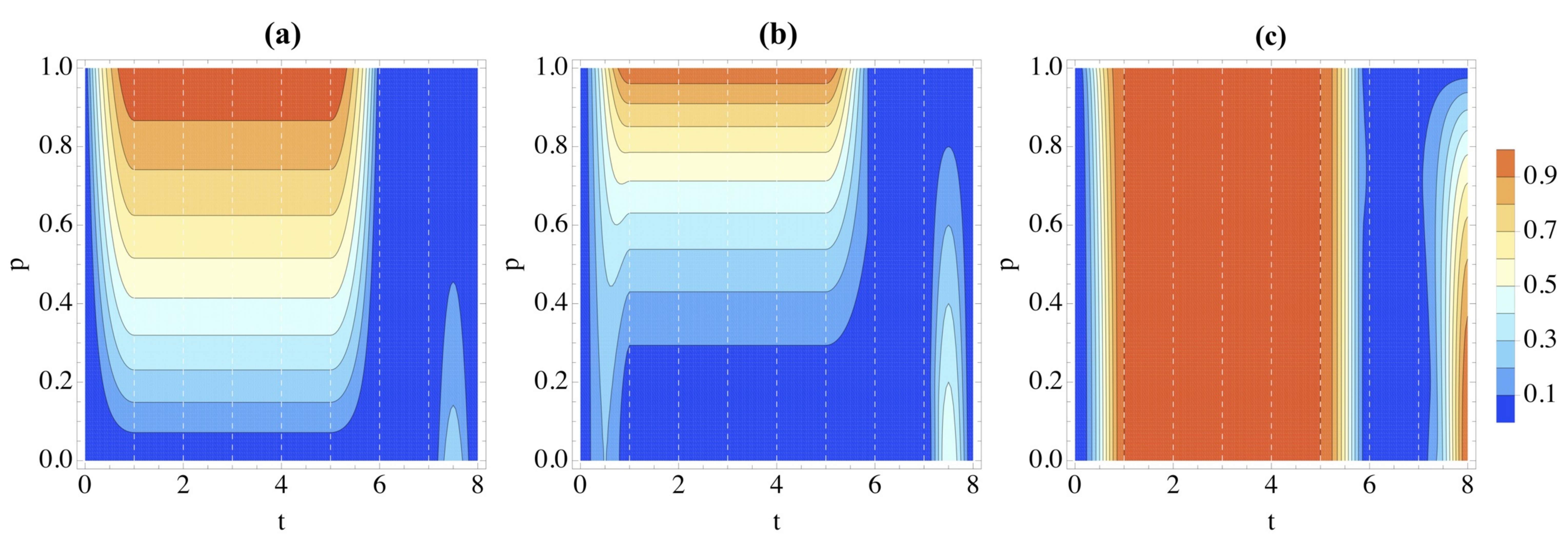}
    \caption{Correlations in the partition $S|E_1 E_2$ as quantified by {\bf (a)} logarithmic negativity, {\bf (b)} discord and {\bf (c)} classical correlations when the Hamiltonian of the system and environment is given by Eq.~\eqref{eq:8Hamiltonian}. We take the initial state of the system to be $\ket{+}$ and the environment $W(p)$.}
    \label{fig:sysEnvCorrDistinctPlus}
\end{figure*}

At first glance, the two types of dynamics seem to result in similar dynamics. However, we see stark changes when we change the system's initial state. Whereas the features of the correlation dynamics remain much the same for the dynamics in Eq.~\eqref{eq:hamiltonian}, they are remarkably different when we change the initial state from $\ket{0}$ to $\ket{+}$ when the Hamiltonian is that in Eq.~\eqref{eq:8Hamiltonian}. This can be easily seen by comparing Figs.~\ref{fig:sysEnvCorrDistinct0} and \ref{fig:sysEnvCorrDistinctPlus}. After the initial CNOT operation, the system and environment become entangled; this is reflected in both Figs.~\ref{fig:sysEnvCorrDistinctPlus} {\bf (a)} and {\bf (b)}. This means that we now see more discord between $S$ and $E_1 E_2$ for larger $p$ rather than smaller; the opposite trend when the initial state is $\ket{0}$. However, entanglement is similar; the more entanglement, the more non-Markovianity. Now we see a small spike during the final gate of the circuit, similar to the unusual resurgence of entanglement when $p=0$ in the overlapping gates case. 

\section{Conclusions} 
\label{conc}
We have critically addressed a measurement-free quantum teleportation protocol against the claim that non-Markovianity is needed in order to boost the teleportation performance~\cite{Tserkis2022}. We have shown that such a connection crucially depends on the way the operations entailed by the protocol are actually implemented. When chopping the circuit in individual gates acting on -- in general multiple -- elements of the register, a more definite relation between the performance of teleportation and non-Markovianity can be spotted, while the evidence of a key role played by non-Markovianity remains weak. On one hand, the negative results reported here reinforce the need for a comprehensive and rigorous resource theory of non-Markovianity for quantum information processing (some interesting initial attempts at establishing such a theory have been made~\cite{Bhattacharya2021,Berk2021}). On the other hand, it emphasises the relevance of the actual way a given quantum protocol is implemented in determining the quantities that are fundamental to its performance. 

\acknowledgments
We thank Spyros Tserkis for invaluable help during the early stages of this work. We acknowledge support by
the European Union’s Horizon Europe EIC-Pathfinder
project QuCoM (101046973), the Royal Society Wolfson Fellowship (RSWF/R3/183013), the UK EPSRC
(EP/T028424/1), and the Department for the Economy
Northern Ireland under the US-Ireland R\&D Partnership Programme.

\bibliography{refs.bib}

\end{document}